\documentclass[twocolumn,floatfix,showpacs,showkeys,superscriptaddress,nofootinbib]{revtex4-1}
\usepackage{graphicx}
\usepackage{amssymb}
\usepackage{amsmath}
\usepackage{bm}
\begin{document}

\setcounter{page}{0}
\title{Dark Energy with  Logarithmic Cosmological Fluid}
\author{Seyen  \surname{Kouwn}}
\email{seyen@skku.edu}
\affiliation{Department of Physics and Institute of Basic Science,
Sungkyunkwan University, Suwon 440-746, Korea}
\affiliation{Institute for the Early Universe, Ewha Womans University, Seoul 120-750, Korea}
\author{Phillial  \surname{Oh}}
\email{ploh@skku.edu}
\affiliation{Department of Physics and Institute of Basic Science,
Sungkyunkwan University, Suwon 440-746, Korea}

\date{\today}

\begin{abstract}
We propose a dark energy model with a logarithmic  cosmological fluid
 which can result in a very small current value of the dark energy density and avoid the coincidence problem without much fine-tuning.
We construct a couple of dynamical models that could realize this dark energy at very low energy in terms of the quintessence of four scalar fields
 and discuss the current
acceleration of the Universe. Numerical values can be made to be consistent with the accelerating Universe
 by adjusting the two parameters of the theory. The potential can be given only in terms of the scale factor, but the explicit form  at very low energy
 can be obtained in terms of  the scalar field to yield of the form $ V(\phi)=\exp(-2\phi)(\frac{4 A}{3}\phi+B)$.
    Some discussions and the physical implications
of this approach are given.
\end{abstract}



\maketitle

\section{Introduction}

One of the most intriguing discovery of modern cosmology is the acceleration of
the Universe \cite{Riess:1998cb,Perlmutter:1998np}
and  dark energy of a repulsive nature is widely believed to be causing the current acceleration.
Many candidates  for the dark energy have been proposed \cite{Carroll:2003qq,Copeland:2006wr,Sahni:2006pa}, among which the cosmological constant is the most accepted one.
 Along with yet another unidentified constituent of the Universe called dark matter, they
compose the standard cosmological model, $\Lambda$CDM \cite{Komatsu:2008hk},
which can address   the observable Universe remarkably well.
Still, an extreme fine-tuning of the cosmological constant \cite{Weinberg:1988cp}  has been the unsatisfactory feature of the model.
That is,  the current cosmological constant must have an unnaturally small value compared to the Planck scale.

An alternative proposal to explain the dark energy is the quintessence model \cite{ratra, caldwell} in which
 a  scalar field  is added as an indispensable component of the Universe. In this approach,
the smallness of the cosmological constant is achieved by a dynamical decay of the
scalar field energy density.
It has the very attractive features of tracking behaviors and attractor solutions \cite{Copeland:2006wr}
 so that galaxy formation is not affected too much by the
quintessence field and the dark energy becomes dominant only at the late stage
of the Universe, thereby causing the current acceleration.
However, in order to achieve the late time dark energy dominance, thus providing a possible
solution of the coincidence problem
\cite{Carroll:2003qq},
the theory has to be fine-tuned to
a certain extent so that the energy density today is very close to the critical density \cite{Copeland:2004cz}.

In this paper, we propose a dark energy model that can alleviate the fine-tuning problem substantially.
Suppose that a  cosmological term \cite{Overduin:1998zv} decaying according to $\Lambda_a\equiv1/a(t)^2$, where $a(t)$ is
the scale factor of the Universe that we regard as the size of the Universe, exists \cite{Ozer:1985ws,Chen:1990jw}. 
This term decreases with the expansion of the Universe, and
the current value is  $\Lambda_a \sim 10^{-122}M_p^2,$ with $a_0\sim 10^{42} ~{\rm Gev}^{-1}$,
 where we have  assumed  its value at the Planck scale to be of the order
$M_p^2.$ Note that the current value of  energy density of this cosmological term  is
very close to the critical density $\rho_{\rm cr,0}\simeq 10^{-122}M_p^4$ and that it has the
potential of explaining the coincidence problem without fine-tuning. Also,  some theoretical background was given
for such a  decay \cite{Ozer:1985ws,Chen:1990jw}.
On the other hand, the conservation of the Einstein tensor prevents
the cosmological term from varying in pure gravity.
 If matter contents are included, the varying cosmological constant
term disrupts the matter continuity equations, which changes the predictions of the standard
cosmology in the matter-dominated epoch \cite{Ozer:1985ws,Chen:1990jw}.
However, if the continuity equation is enforced, the $1/a^2$ term
  behaves exactly like the curvature constant term and it alone cannot yield an accelerating Universe

It turns out that by adding another cosmological term that  varies according to  $\ln a/a^2$ to the previous $\Lambda_a$, an accelerating Universe can be realized.
To see this in detail, 
we first assume that the energy density of the cosmological fluid composed of these two terms 
 is approximately of the order of $\sim {\rm M_{p}}^4$ when the inflation is started{\footnote{The cosmological terms considered here decay very fast and cannot be responsible for the inflation
itself. The inflation should come from some other source.}}.
At the end of the inflation, when the scale factor becomes $10^3$ cm with the number of $e$-foldings given by $N$,
 the energy density will have decreased in magnitude by
a factor of $\sim Ne^{-2N}$ and will have become of the order of $\sim N e^{-2N}M_p^4$.
We choose $N \sim 81${\footnote{This number may look rather ad-hoc, but the choice of the  e-foldings is related with the scale factor at the end of inflation and can be adjusted.}}, which is bigger than the minimum number of e-foldings \cite{Copeland:2004cz}
so that the energy density becomes of the order of $ \sim 10^{-72}{\rm M_{p}}^4$. Therefore, we propose the following dark energy density:
\begin{eqnarray}
\rho_{D} =\left[ \frac{c_{*} \ln (a/a_{{\rm inf}})}{ (a/a_{{\rm inf}})^2}
+\frac{d_{*}}{ (a/a_{{\rm inf}})^2}\right ]
\times 10^{-72}{\rm M_{p}}^4,\label{ded}
\end{eqnarray}
where $c_*$ and $d_*$ are constants{\footnote {We also assume that the pressure is given by
$p_{D} =-\frac{1}{3}\left[ \frac{c_{*} \ln (a/a_{{\rm inf}})}{(a/a_{{\rm inf}})^2}
+\frac{c_* + d_{*}}{ (a/a_{{\rm inf}})^2}\right ]
\times 10^{-72}{\rm M_{p}}^4.$ $\rho_{D}$ and $p_{D}$ satisfy the continuity equation, $\dot\rho_{D}= -3H(\rho_{D}+ p_{D})$.}}
 and $a_{{\rm inf}}\sim 10^3 {\rm cm}$.  Note that the number $10^{-72}$ is a dynamical consequence of the inflation, and no extreme fine-tunings of $c_*$ and $d_*$ turn out to be necessary to
describe the current accelerating Universe. 
The current observations give $\rho_{D,0} = \delta \rho_{cr},~\delta \simeq 0.73$  with $\rho_{cr} \simeq 10^{-122} {\rm M_{p}}^4$,  and the matter energy density $\rho_m = {\gamma  \over (a/a_0)^3}\rho_{cr},~\gamma \simeq 0.27$ \cite{Komatsu:2008hk}. Then,  Eq. (\ref{ded}) yields  a relation
 \begin{eqnarray}
 c_*( 25\ln 10) + d_* \sim \delta.       \label{ded1}
\end{eqnarray}

Just after the inflation ends at the energy scale $\sim 10^{13}~ {\rm Gev}$, the energy
density is of the order of $10^{-24}M_p^4>> \rho_{D}$, and
$\rho_{\rm r}\sim 10^{-24} / (a/a_{r,i})^4M_p^4 $
with $a_{r,i}\sim a_{{\rm inf}} \sim 10^3 {\rm cm} $.
The normal expansion takes over and the Universe expands by a factor of $10^{21}$ until
the radiation-matter equality around $a\sim 10^{24} {\rm cm}$.
When this occurs, $\rho_{r,f}\simeq \rho_{m,i} \simeq 10^{-108}{\rm M_{p}}^4$, and  the matter dominance takes over because the radiation energy density
decays faster than the matter energy density. In the meantime,
the magnitude of $\rho_{D}$ keeps on decreasing according to Eq. (\ref{ded}) and becomes of the order of $\sim 10^{-114}{\rm M_{p}}^4.$
Therefore, the dark energy is completely subdominant during this period.
Then, the matter-dominated epoch begins around $a \simeq 10^{-3}a_0,~a_0\sim 10^{28} {\rm cm}$.
Because the matter energy density decays faster than dark energy,  a scale where $\rho_{m} \simeq \rho_{D}$ exists and is given by
 \begin{eqnarray}
 \frac{a_{{\rm eq}}}{ a_0}\left[ c_* \ln\left({a_{eq}\over a_0}\right) + \delta \right]\simeq \gamma. \label{ded2}
 \end{eqnarray}

So far, there is only one restriction on the numerical values of the parameters $c_*$ and $d_*$ of Eq. (\ref{ded1}), and a wide range of their values are allowed to fit into the current observation.
 One can impose one more condition by demanding that the Universe began its acceleration very recently. The acceleration equation with our $\rho_{D}$ and $p_{D}$
is given by
  \begin{eqnarray}
   {\ddot{a}\over a} = \frac{1}{6}\left[\frac{c_*}{(a/a_0)^2} -
   \frac{\gamma}{ (a/a_0)^3}\right]\rho_{cr}.\label{acceqmo}
\end{eqnarray}
From the above equation, we see that the acceleration began around $a_{{\rm acc}} \sim \gamma/c_* a_0$. For example, if we
 choose $c_* = 0.54$, we obtain $a_{{\rm acc}}\sim 1/2 a_0$ and $a_{{\rm eq}} \sim 0.60 a_0$. This would determine $ d_*\sim -56.$  Therefore, the transition to dark energy dominance occurred very recently.
With these values,  the dark energy stays negative during
 most of the time until $a_{-+}\sim 0.25 a_0$
and becomes positive at a late stage of
the matter-dominated era. After passing this point,
a maximum, $a_{{\rm max}}$ is reached,
and eventually, it begins to  be dominant
around $a_{{\rm eq}}$.
It seems that a priori there is no reason for the dark energy to stay positive always as long as
the total energy density $\rho=\rho_{D}+\rho_r$ or $\rho=\rho_{D}+\rho_m$ remains positive.
In addition, the absolute value of the energy density is very small when it stays negative compared to the radiation or matter energy density.
 Therefore, it should
not disturb the radiation- or matter-dominated evolution  altering the course of it {\footnote{
One can check that $a_{-+}\sim e^{-\delta/c_*} a_0$ independently of the
detailed numerical values for $\rho_{{\rm cr}}$ and
$a_{{\rm inf}}$.}}.
Note that the increasing behavior of dark energy, although limited until $a_{{\rm max}}$ in our case, also appears in the phantom model \cite{Caldwell:1999ew}.
It is interesting to see that $a_{-+}, a_{{\rm max}},
a_{{\rm acc}},$ and $ a_{{\rm eq}}$ all happen very recently
without too much fine-tuning of the parameters
$c_*$ and $d_*$.

In the next sections, we  show that Eq. (\ref{ded})
can be realized within a couple of  quintessence models at very low energy.
In Section II, we construct a quintessence model
with four scalar fields that can produce
the dark energy behavior of Eq. (\ref{ded}). An explicit form of the potential
in terms of the scalar field in the scalar-dominated region is given. In Section III, we consider a generalized quintessence model
where the explicit construction can be extended to the matter-dominated epoch. In Section. IV, a critical analysis
of the generalized quintessence model is performed. Section V contains conclusions and discussions.


\section{Quintessence with four scalar fields}

Let us consider an action of the form ($8\pi G=1$)
\begin{eqnarray}
S_1 = \int d^4 x \sqrt{-g} \biggr[{R \over 2}-{1\over2}\partial_\mu \phi \partial^\mu \phi -V(\phi) -\alpha X  \biggr] + S_m \,, \nonumber
\end{eqnarray}
with
\begin{eqnarray}
X \equiv g^{\mu\nu}\delta_{ab}\partial_\mu \sigma^a\partial_\nu \sigma^b \,,
\end{eqnarray}
where $\sigma^a$ is  3-component scalar field and $\alpha$ is a parameter that we assume to be constant. $S_m$ ia a matter action with $p_m=0.$
The space-time metric tensor is given by
\begin{eqnarray}
ds^2 = -dt^2 + a(t)^2(dx^2 + dy^2 + dz^2) \,,
\end{eqnarray}
where $a(t)$ is the scale factor.
With an ansatz for the scalar field $\sigma^a$ of the form $\sigma^a = x^a$ \cite{Lee:2009zv},
the evolution and the continuity equations for matter are given as  follows:
\begin{eqnarray}
3H^2 &=&  {1 \over 2} \dot{\phi}^2 + { 3 \alpha \over a^2 }  + V + \rho_m \,, \label{ScFEqA} \\
-2 \dot{H} &=&  \dot{\phi}^2 + { 2 \alpha \over a^2 }   + \rho_m \,, \label{ScFEqB}  \\
0 &=& \dot{\rho}_m + 3H \rho_m \label{ScFEqC} \,,
\end{eqnarray}
where $H\equiv \dot{a}/a$ is the Hubble parameter, and $\rho_m$ is the matter energy density.
The scalar field satisfies
\begin{eqnarray}
\ddot{\phi} + 3H\dot{\phi}  + V'(\phi) = 0 \,, \nonumber
\end{eqnarray}
where the dot and the prime denote partial differentiations with respect to $t$ and $\phi$, respectively.

Taking linear combinations of Eqs. (\ref{ScFEqA}) and (\ref{ScFEqB}), we get the acceleration equation
\begin{eqnarray}
{\ddot{a}\over a} &=& {1\over3}\left(V-\dot{\phi}^2 \right) - {1\over6}\rho_m \,. \label{ScTenAccRate}
\end{eqnarray}
In the quintessence model, the kinetic term and the matter density decay very fast
in the above equation, and sole potential dominance at late time is reached, which causes the acceleration.
We open the possibility that the late time acceleration comes from the first two terms in the
above equation, and
anticipating that Eq. (\ref{ScTenAccRate}) will describe the current acceleration at very low energy,
we require the following relation:
\begin{eqnarray}
V-\dot{\phi}^2
= {A \over (a/a_0)^2}\rho_{{\rm cr}}  \,, \label{OurChooseAnsantz}
\end{eqnarray}
where
$A$ is a parameter that we assume to be the positive.
 With this, the expansion changes from
a deceleration for a small value of the scale factor corresponding
to the matter-dominated epoch to an acceleration for a large value of the scale factor corresponding
to the scalar-dominated epoch, including the kinetic energy density. We will omit $a_0$ and $\rho_{{\rm cr}}$ in what follows
unless confusion arises.
Using the relation in Eq. (\ref{OurChooseAnsantz}), we can rewrite the evolution equations, Eqs. (\ref{ScFEqA}) and (\ref{ScFEqB}) as follows:
\begin{eqnarray}
3H^2 &=& {3\over2}V - {A\over 2a^2} + {3\alpha \over a^2} + \rho_m  \,, \label{ScFEqAA} \\
-2 \dot{H} &=&  V - {A \over a^2} + { 2 \alpha \over a^2 } + \rho_m  \,. \label{ScFEqBB}
\end{eqnarray}
Differentiating Eq. (\ref{ScFEqAA}) with respect to time and comparing the result with Eq. (\ref{ScFEqBB}),
we obtain the following first-order differential equation for the potential:
\begin{eqnarray}
3aV'(a)+6V(a)-{4A\over a^2} = 0 \,, \label{ScFEqDiffPoten}
\end{eqnarray}
which yields
\begin{eqnarray}
V =  {4A \ln a \over 3a^2} + {B \over a^2} \,, \label{ScFSolVa}
\end{eqnarray}
which is precisely of the form in Eq.(\ref{ded}). From Eq. (\ref{ScFEqAA}), we define (with $\alpha=-A/3$, see Eq. (\ref{messia}) and below)
\begin{eqnarray}
\rho_{\phi} = {2A \ln a \over a^2} + {3(B-A) \over 2 a^2}.\label{energyded}
\end{eqnarray}
Comparing with Eq. (\ref{ded}), we obtain $A=c_*/2, ~3B=3c_*/2 + 2\delta.$

Note that it is difficult to express the scalar field $\phi$ in a closed form from Eq. (\ref{OurChooseAnsantz}),
\begin{eqnarray}
\phi = \int dt \sqrt{ {4A\ln a\over3a^2}+{B-A\over a^2} },
\end{eqnarray}
so the potential $V$ cannot be expressed in terms of the $\phi$ explicitly.
Also, the positivity of the square root in the above expression restricts the applicable range of $a$, which turns out to be $a\gtrsim a_{-+}$, and this  is the same as the positivity of the $\rho_{\phi}$ of Eq. (\ref{energyded}).
Basically, this  comes from the imposition of the acceleration condition, Eq. (\ref{OurChooseAnsantz}), and
 this condition restricts the dynamically permitted region to
$a \geq a_{-+}.$
A closed form of the potential is viable, if we neglect
the matter density in the evolution equations, that is, in the scalar-dominated epoch.
In this case,
the ratio of $\dot{\phi}^2$ to $H^2$ is given by
\begin{eqnarray}
{\dot{\phi}^2 \over 3H^2} = {  V-{A\over a^2} \over {3\over2}V - {A\over 2a^2} + {3\alpha \over a^2} } \,,\label{messia}
\end{eqnarray}
and we can adjust the parameter $\alpha = -{1\over3}A$. Then, we have
\begin{eqnarray}
\phi(a) = \sqrt{2} \ln a + C\,, \label{ScFSolPhia}
\end{eqnarray}
with an integration constant $C$.
 This gives an expression for the potential in Eq. (\ref{ScFSolVa}):
\begin{eqnarray}
V(\phi) &=& e^{-{\sqrt 2}\phi} \left( {2{\sqrt 2}A^\prime\over 3}\phi 
+ B^\prime  \right),\\ 
 A^{\prime}&=&e^{\sqrt{2}C}A,~
  B^\prime=B-\frac{2\sqrt{3}}{3}AC\,\nonumber\label{ScTenSolVSCDom}.
\end{eqnarray}

\section{Generalized quintessence model}
In this section, we consider a generalized quintessence model given by
\begin{eqnarray}
S_2 = \int d^4 x \sqrt{-g} \biggr[{R \over 2}-{\omega(\phi)\over2}\partial_\mu \phi \partial^\mu \phi -V(\phi)
-\alpha X  \biggr] + S_m \,,~~~~~ \label{gqm}
\end{eqnarray}
 where we have introduced a scalar function $\omega(\phi)$, which will be
 arranged for our purpose. The last term in the gravity sector is the triplet of scalar fields as before,
 and we choose the same ansatz of the form $\sigma^a = x^a$, which solves the equation of motion.
  The potential $V(\phi)$ takes the following form:
\begin{eqnarray}
V(\phi) &=& e^{-2\phi} \left( {4A\over 3}\phi + B  \right) \label{ScTenSolVPhi} \,,
\end{eqnarray}
which is essentially of  the form given by Eq. (\ref{ScTenSolVSCDom}) with a redefinition of the scalar field.
We assume that the matter is cold dark matter with $w_m=0$ as before and that it satisfies the continuity equation.
The evolution equations  are given by
\begin{eqnarray}
3H^2 &=& {1\over2}\omega\dot{\phi}^2 + { 3 \alpha \over a^2 } + V + \rho_m  \,, \label{ScTenEqA} \\
-2\dot{H} &=& \omega\dot{\phi}^2 + { 2 \alpha \over a^2 } +\rho_m  \,, \label{ScTenEqB}
\end{eqnarray}
along with the continuity equation, Eq.  (\ref{ScFEqC}),
and the scalar field satisfies
\begin{eqnarray}
\omega\ddot{\phi}+{1\over2}\omega'\dot{\phi}^2+3 H \dot{\phi} + V'  = 0 \,. \label{ScTenEqScalar}
\end{eqnarray}

Taking linear combinations of Eqs. (\ref{ScTenEqA}) and  (\ref{ScTenEqB}),
 we get the acceleration equation
\begin{eqnarray}
{\ddot{a}\over a} &=& {1\over3}\left(V-\omega\dot{\phi}^2\right) -\rho_m \,. \label{ScTenEqAcc}
\end{eqnarray}
   We choose the same acceleration condition as before,
\begin{eqnarray}
V-\omega\dot{\phi}^2 = {A \over a^2}  \,, \label{ScTenChoose}
\end{eqnarray}
with a positive $A$. Note that Eqs. (\ref{ScTenEqA}) and (\ref{ScTenEqB}) imply Eq.
(\ref{ScTenEqScalar}) and recall $\rho_m={\gamma  \over (a/a_0)^3}\rho_{cr}$.
Therefore, we have two independent dynamical equations and one constraint, Eq. (\ref{ScTenChoose}), 
for the three unknown functions to be determined, $a(t)$ or $H$, $\omega(\phi)$, and $\phi(t)$.
It turns out that we can solve the equations with the ansatz
\begin{eqnarray}
\phi(a) =  \ln a   \,,    \label{ScTenAnsatz}
\end{eqnarray}
which reproduces the same forms of the scale-factor-dependent potential,
Eq.  (\ref{ScFSolVa}), and the energy density, Eq. (\ref{energyded}).
Inserting this ansatz into Eq. (\ref{ScTenChoose}), we obtain the following $\omega$ in terms of $\phi$:
\begin{eqnarray}
\omega(\phi) &=& {8A\phi + 6(B-A)  \over 4A\phi +3(B-A) + 2\gamma e^{-\phi}    }     \,. \label{ScTenSolOmegaPhi}
\end{eqnarray}
Note that for the choice of $\alpha=-A/3$, $\omega(\phi)> 0$ for a positive
value of the energy density $\rho_{\phi}$ defined with the first three terms of Eq. (\ref{ScTenEqA})
as before.
One can show that for $\alpha<-A/3$, $\omega(\phi)$ is always greater than zero for $a \gtrsim a_{-+}.$
The above analysis shows that it is possible to construct an explicit form of the potential beyond the scalar-dominated region if
we introduce a generalized quintessence model with an adjustable kinetic function $\omega(\phi).$

\section{Critical Analysis}
In this section, we perform a critical analysis \cite{Copeland:2006wr} of the evolution equations of the previous section
and present some numerical result. For convenience, we choose $\alpha=-A$. For other choices, the qualitative feature
of the stability does not change as long as  $\alpha<-A/3$.
Let us introduce the following dimensionless quantities:
\begin{eqnarray}
x\equiv\frac{\omega\dot{\phi}^2}{6H^2} \,,~
y\equiv\frac{\tilde V}{3H^2} \,,
\end{eqnarray}
with $\tilde V = V+ 3\alpha/a^2.$
Then, Eqs. (\ref{ScTenEqA})-(\ref{ScTenEqScalar}) can be written in the following form:
\begin{eqnarray}
\frac{dx}{dN} &=&  -3x^2 +{5\over3}x  -{1\over3}y \,, \label{CriticalDx} \\
\frac{dy}{dN} &=&  {1\over3}y - 3xy + {4\over3}x  \,, \label{CriticalDy}
\end{eqnarray}
where 
$N\equiv \ln a$,  together with a constraint equation
\begin{eqnarray}
\frac{\rho_m}{3H^2} = 1-x-y \,. \label{CriticalConstraint}
\end{eqnarray}
The critical points of the above system are easily obtained by setting the right-hand sides of the above equations, Eqs. (\ref{CriticalDx}) and (\ref{CriticalDy}), to zero.
The only physically meaningful critical points $(x_c,y_c)$ of the system are
\begin{eqnarray}
{\rm (A)}:&&~ (x_c, y_c) = (0,0) \,, \nonumber \\
{\rm (B)}:&&~ (x_c, y_c) = \left({1\over3},{2\over3}\right) \,. \nonumber
\end{eqnarray}
Point (A) corresponds to the matter-dominated point whereas point (B)
is the scalar-dominates one.

To gain some insight into the property of the critical points, we write the variables near the critical points $(x_c,y_c)$
in the forms\begin{eqnarray}
x = x_c + \delta x, \\
y = y_c + \delta y,
\end{eqnarray}
where $ \delta x$ and $ \delta y$ are perturbations around the critical points.
From Eqs. (\ref{CriticalDx}) and (\ref{CriticalDy}), we obtain the linearized equations
\begin{eqnarray}
\frac{d\delta x}{dN}  &=&  \left({5\over3}-6x_c \right)\delta x  -{1\over3} \delta y \,,\label{CriticalDX} \\
\frac{d\delta y}{dN}  &=& \left({4\over3}-3y_c \right)\delta x  + \left({1\over3}-3x_c \right)\delta y  \,,\label{CriticalDY}
\end{eqnarray}
which can be written by using a matrix $M$ as
\begin{eqnarray}
 {d\over d N}
 \left(
  \begin{array}{c}
    \delta x \\
    \delta y \\
  \end{array}
\right) = M
 \left(
  \begin{array}{c}
    \delta x \\
    \delta y \\
  \end{array}
\right) \,,~ M =
\left(
  \begin{array}{cc}
     {5\over3}-6x_c   &  -{1\over3} \\
     {4\over3}-3y_c  &  {1\over3}-3x_c \\
  \end{array}
\right) \,.~~~~~~ \label{CriticalMatrix}
\end{eqnarray}
One can study the stability of the critical points against perturbations by evaluating the eigenvalues of the matrix $M$.
For class $(\rm A)$ corresponding to the matter-dominated epoch,  $\lambda_1=1$ and $\lambda_2=1$, which means that it is an unstable point.
For class $(\rm B)$ corresponding to the scalar-field-dominated epoch,  $\lambda_1=-1$ and $\lambda_2=0$. The appearance of a zero eigenvalue means that
the linear perturbation, which leads to the matrix in Eq. (\ref{CriticalMatrix}), is not adequate, so the higher-order
perturbations must be considered to determine whether the considered critical point is stable or not.

First, note that we have
\begin{eqnarray}
2x-y > 0\,,
\end{eqnarray}
for $\alpha=-A.$
Therefore, it is useful to change from  $(x,y)$ to the new variables $(X,Y)$ defined as
\begin{eqnarray}
X = x + 2y \,,~ Y= 2x - y \,,
\end{eqnarray}
where the $X$, $Y$ are positively defined. Then,  Eqs.  (\ref{CriticalDx}) and (\ref{CriticalDy}) can be rewritten as
\begin{eqnarray}
\frac{dX}{dN} &=&  {5\over3}Y -{1\over5}X(3X+6Y-5) \,, \label{CriticalCheckX} \\
\frac{dY}{dN} &=&  -{1\over5}Y(3X+6Y-5)  \,, \label{CriticalCheckY}
\end{eqnarray}
and the corresponding critical points in terms of  the new variables are
\begin{eqnarray}
{\rm (A)}: &&~ (X_c,Y_c)=(0,0) \,, \nonumber \\
{\rm (B)}: &&~ (X_c,Y_c)=\left({5\over3},0\right)\,. \nonumber
\end{eqnarray}

In the higher-order perturbations, we write the variables near the critical points $(X_c,Y_c)$ in the forms
\begin{eqnarray}
X &=& X_c + \delta X^{(1)} + \delta X^{(2)} + \delta X^{(3)} + \cdots \,, \nonumber \\
Y &=& Y_c + \delta Y^{(1)} + \delta Y^{(2)} + \delta Y^{(3)} + \cdots \,,
\end{eqnarray}
where $\delta X^{(n)}$ and $\delta Y^{(n)}$ are $n$-th order perturbations of the variables near the critical points.
For class $(\rm B)$, the perturbative equations of each order are given by
\begin{eqnarray}
{d \over dN} \, \delta X^{(1)} &=& -\delta X^{(1)} -{1\over3}\delta Y^{(1)  }, \nonumber \\
{d \over dN} \, \delta Y^{(1)} &=& 0,  \nonumber \\
{d \over dN} \, \delta X^{(2)} &=& -{3\over5}\delta X^{(1)} \left( \delta X^{(1)} +2\delta Y^{(1)} \right)
                                                 -\delta X^{(2)} -{1\over3}\delta Y^{(2)}, \nonumber \\
{d \over dN} \, \delta Y^{(2)} &=& -{3\over5}\delta Y^{(1)} \left( \delta X^{(1)} +2\delta Y^{(1)} \right),  \nonumber \\
{d \over dN} \, \delta X^{(3)} &=& -{6\over5}\delta X^{(1)}\delta X^{(2)}
                                                 -{6\over5}\left(\delta X^{(1)}\delta Y^{(2)}+\delta X^{(2)}\delta Y^{(1)}\right) \nonumber \\
&&-\delta X^{(3)} -{1\over3}\delta Y^{(3)}, \nonumber \\
{d \over dN} \, \delta Y^{(3)} &=&-{12\over5}\delta Y^{(1)}\delta Y^{(2)}
                                                 -{3\over5}\left(\delta X^{(1)}\delta Y^{(2)}+\delta X^{(2)}\delta Y^{(1)}\right), \nonumber \\
&\vdots& \nonumber
\end{eqnarray}
Note that the right-hand side of the second equation in the first-order perturbation equation is zero, which reflects the fact that
one of two eigenvalues is zero, so we must focus on the next-order equations.

The solutions of the  above linear differential equations are given by
\begin{eqnarray}
X(N) &=& {5\over3} + \delta A_1 + \delta B_1  \,, \\
Y(N) &=&  0 +  \delta A_2 +\delta  B_2   \,,
\end{eqnarray}
with
\begin{eqnarray}
\delta A_1 &=& -{1\over3}\delta Y_0 +{1\over3}\delta Y_0^2 N -{1\over3}\delta Y_0^3 N^2 + {\cal O}(\epsilon^4), \\
\delta A_2 &=& \delta Y_0 -\delta Y_0^2 N +\delta Y_0^3 N^2 + {\cal O}(\epsilon^4), \\
\delta  B_1 &=& \delta\alpha_0 e^{-N}\Biggr[ 1+ {3\over5}e^{-N}\delta\alpha_0 -\left({1\over5}+N\right)\delta Y_0 \nonumber \\
   &&~~~~~~~~~~~+{9\over25}e^{-2N}\delta\alpha_0^2 -{3\over25}e^{-N}\left(1+10N\right)\delta \alpha_0 \delta  Y_0 \nonumber \\
   &&~~~~~~~~~~~ +N\left(1+{2\over5}\right)\delta Y_0^2 \Biggr] + {\cal O}(\epsilon^4) \,, \\
\delta B_2 &=& {3\over5}\delta  Y_0 \delta\alpha_0 e^{-N}\Biggr( 1+ {3\over5}e^{-N}\delta\alpha_0 -2N \delta Y_0 \Biggr) + {\cal O}(\epsilon^4)\,, \nonumber 
\end{eqnarray}
where $\delta Y_0$ and $\delta\alpha_0$ are the initial values at $N=0$
that satisfy    $\delta Y_0 = \delta Y^{(1)}(0)$, and $\delta\alpha_0 = \delta X^{(1)}(0)+{1\over3}\delta Y^{(1)}(0)$, respectively,  and
$\epsilon$ is the infinitesimal order parameter for the perturbation.
 When the higher-order terms are included and  when $N$ becomes very large, $\delta A_{1,2}$
 can be expressed in a closed form with
\begin{eqnarray}
\delta A_{1} &\rightarrow &-{\delta Y_0\over3(1+\delta Y_0 N)} \,,~ \delta B_{1}\rightarrow 0 \,, \\
\delta A_{2} &\rightarrow&{\delta Y_0\over1+\delta Y_0 N} \,, \,~~~~~~~\delta B_{2} \rightarrow 0 \,.
\end{eqnarray}
Because the variable $Y$ is a
positively-defined value, the perturbation around the zero point must  also be positive.
In this case, $\delta A_1$ and $\delta A_2$ smoothly go to zero
 when $N$ goes to infinity. Therefore, the critical point (B)
is stable.
 The numerical result is given in Fig. 1,  which demonstrates the
stability of the scalar-dominated critical point.

\begin{figure}[ht]
\begin{center}
\scalebox{0.5}[0.45]{ \includegraphics{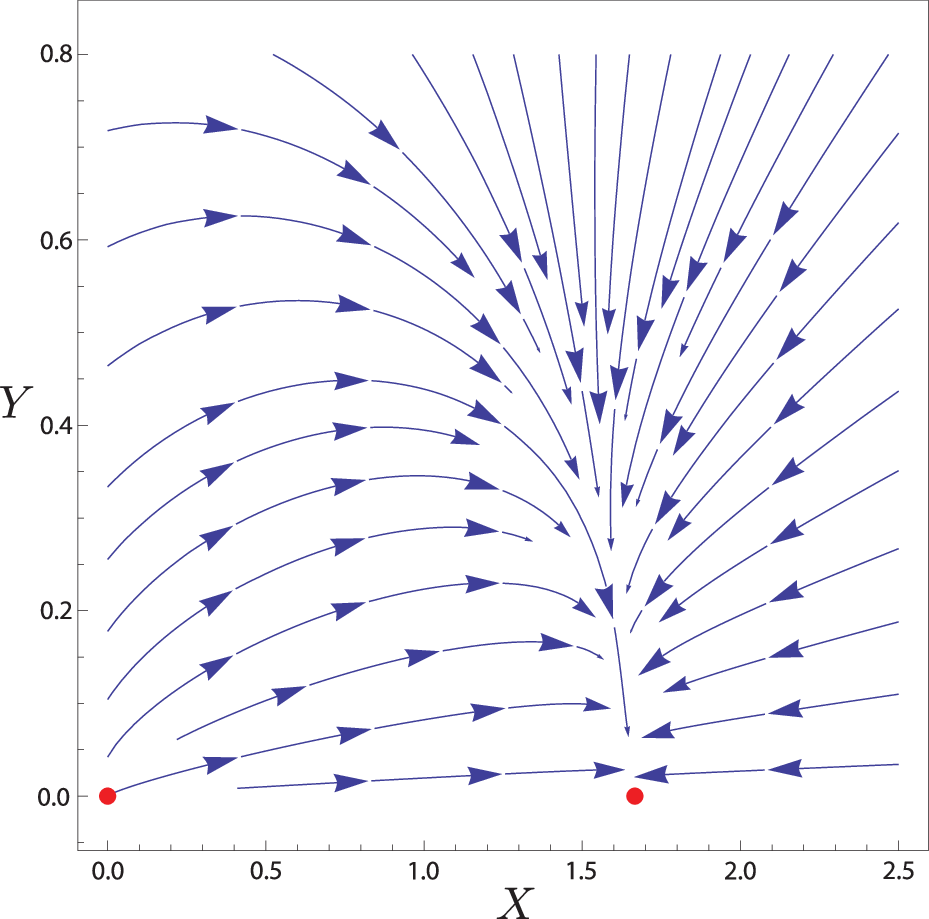}}
\end{center}
\caption{\small Flow diagram of the system in terms of $X,Y$ for different initial values.
We have drawn the physically-allowed region with $\delta Y>0$. The left red dot is the matter-dominated point (A)
whereas the right red dot is the scalar-dominated point (B).}
\label{fig000}
\end{figure}

\section{Conclusion}

We showed that the coincidence problem can be avoided with a logarithmic cosmological fluid
of the form in Eq. (\ref{ded}), and this can be realized dynamically as a couple of quintessence models
at very low energy.
Among the four scalar fields, one  plays a major role as in the standard quintessence model, thus
causing the current acceleration. On the other hand,  the triplet of scalar fields is not essential
 in the dynamical evolution,  but it can
provide the necessary energy density such that the potential is completely
integrable as an exact expression  in the dark-energy-dominated era.
In the generalized quintessence model, the construction was extended to
the matter-dominated epoch, and a critical analysis indicates that the scalar domination is dynamically stable.
An analytic
expression of the potential in terms of the scalar field is unavailable in each case,
but  effective field theories  can
be considered  separately at each stage of the evolution of the Universe.
We also have checked that in such a scheme, the
current value of the extremely small dark energy density can be obtained without much
fine-tuning; The constants $c_*$ and $d_*$ are only of the order $10^{-1}$ and $10^2$,
and the small value is attributed  to   decaying of the dark energy density  essentially as $1/a^2$.

Compared with the quintessence model, the constraint in Eq. (\ref{OurChooseAnsantz}) brings a crucial difference as far as energy dominance is concerned. When the scalar field begins to roll down the potential, the initial potential
energy is converted into kinetic energy, which dominates the energy of the scalar field.
However, the kinetic energy decreases rapidly, and  potential energy dominance takes over around
$1/2 <z<1$, which is responsible for the domination of the dark energy. In our case, at  very low energy,  Eq. (\ref{OurChooseAnsantz}) applies
throughout the evolution, which implies that kinetic energy does not decay fast, but  remains comparable to the  potential energy $\sim V/2$.
This would result in a relatively small absolute value for the equation of state $\omega_{\phi}$,
but a detailed comparison with the observational data needs to be done.
We comment that this constraint is a phenomenological input to conform to the current
acceleration, so a more theoretical basis is required.

Another comment is the feature that the dark energy density remains negative during
most of the time until $a_{-+}$. Even though this behavior does not destroy the accelerating Universe, $a \sim 0.25 a_0$ seems to have no special meaning
  in the evolution.
 How the matter-dominated evolution is affected by the small
negative energy, if at all, needs to be addressed.  One could get rid of the negative energy simply by modifying the dark energy density to 
$\rho_{D}=-{\rm RHS}$ of Eq. (\ref{ded}) for $a>a_{-+}$,
but this seems  rather ad-hoc.
If this scale can somehow be raised to electro-weak breaking scale, this  would  endow the theory with more flexibility.
Adding $(\ln a)^2/a^2$ to the dark energy density in Eq. (\ref{ded})  is another possible avenue to deal with the negative energy.
It is likely that this will also extend the applicable range of the quintessence model similarly constructed and
might  improve the equation of state previously mentioned.

We conclude with an intriguing  property of Eq. (\ref{ded}). One can check that
that the status of the current  accelerating Universe is rather insensitive
to the numerical values of $a_{\rm inf}\sim 10^3 cm$ and $\rho_{{\rm cr}}\sim 10^{-122}M_p^4$
 chosen. If these values are chosen differently, these changes
can  always be reabsorbed into $c_*$ and $d_*$. Suppose we change $\rho_{{\rm cr}}$
to $\alpha^{-1} \rho_{{\rm cr}}$ with some constant $\alpha$.
Then, Eqs. (\ref{ded}) and (\ref{acceqmo}) retain the same form with
$c_*\rightarrow \alpha c_*$, and $d_*\rightarrow \alpha d_*$.
Likewise, if $a_{{\rm inf}}\rightarrow \beta^{-1} a_{{\rm inf}}$, then, $c_*\rightarrow \beta^{2} c_*$, and $d_*\rightarrow \beta^{2} (
c_* \ln\beta + d_*)$.
Because $\alpha$ and $\beta$ can be
at most of the order 1, $c_*$ and $d_*$ can change  at most by an order of 2, so the fine-tuning problem does not arise.
These changes also do not alter the energy dominance  of the radiation epoch and the matter epoch.
This especially means that  $a_{acc}$ can be made, in fact, independent of $a_{{\rm inf}}$ without much fine-tuning,
suggesting that  the current status of the Universe
is not affected too much by the early Universe.
Similarly, the scaling of the scale factor itself can also be absorbed into
new definitions of $c_*$ and $d_*$, which seems to suggest a kind of scaling behavior of the dark energy proposed.
It would be interesting if some theoretical foundation could be given for this.

\vspace{0.5cm}
~~~~~~~~~~~~~~~ACKNOWLEDGMENTS
\vspace{0.2cm}

We acknowledge the hospitality at APCTP where part of this work was done.
This work
was supported by the National Research Foundation of Korea (NRF) grant funded by the BSRP through the MEST (2010-0021996) and by the
Korea government (MSIP) through the Center for Quantum Spacetime (CQUeST) of Sogang
University with grant number 2005-0049409.


\end{document}